\documentclass[a4paper,11pt]{article}
\usepackage{authblk}
\usepackage{graphicx}
\usepackage{geometry}
\usepackage{longtable}
\usepackage{array}
\usepackage{wrapfig}
\usepackage{braket}
\usepackage[english]{babel}
\usepackage{bbold} 
\usepackage{fancyhdr}
\usepackage{amsmath}
\usepackage{amssymb}
\usepackage[T1]{fontenc}
\usepackage{lmodern}
\usepackage{microtype}
\usepackage[utf8]{inputenc}
\usepackage{eso-pic}
\usepackage{tensor}
\usepackage{empheq}
\usepackage{slashed}
\usepackage{cancel}
\usepackage{bigints}
\usepackage{dsfont}
\usepackage{systeme}
\usepackage[pdfpagelabels]{hyperref}
\usepackage{comment}
\usepackage{authblk}
\usepackage{dirtytalk}
\usepackage[sorting=none]{biblatex}
\hypersetup{colorlinks=true, linkcolor=blue, urlcolor=blue}
\numberwithin {equation} {section}
\title{Experimental observation of hyperbolic spacetime dynamics}

\author[1]{Jonas Himmel$^\ast$}
\author[2]{Coraline Bacq $^\ast$}
\author[3]{Krishna Chand Maurya}
\author[4]{Max Ehrhardt}
\author[1]{Matthias Heinrich}
\author[1]{Tom A. W. Wolterink}
\author[2]{Pablo Basteiro} 
\author[2,5,6]{Rathindra Nath Das}
\author[2,7]{Ren\'{e} Meyer}
\author[3]{Tobias Huber-Loyola}
\author[3]{Andreas Pfenning}
\author[3]{Sven H\"{o}fling}
\author[2]{Johanna Erdmenger}
\author[1]{Alexander Szameit}

\affil[1]{Institute of Physics, University of Rostock, Germany}
\affil[2]{Institute for Theoretical Physics and Astrophysics and W\"{u}rzburg-Dresden Cluster of Excellence ctd.qmat, Julius-Maximilians-Universit\"{a}t W\"{u}rzburg }
\affil[3]{Physikalisches Institut and W\"{u}rzburg-Dresden Cluster of Excellence ctd.qmat, Julius-Maximilians-Universit\"{a}t W\"{u}rzburg, Germany}
\affil[4]{Department of Physics, McGill University, 3600 Rue University, H3A 2T8 Montr\'{e}al, Canada}
\affil[5]{Weizmann Institute of Science, Rehovot, Israel}
\affil[6]{MIT Center for Theoretical Physics-a Leinweber Institute, Massachusetts Institute of Technology,Cambridge, MA 02139, USA}
\affil[7]{Shanghai Institute for Mathematics and Interdisciplinary Sciences (SIMIS),
Shanghai, 200433, China}

\date{}
\newcommand{\arccot}{\operatorname{arccot}}

\newcommand{\pr}[1]{\left(#1\right)}
\newcommand{\pq}[1]{\left[#1\right]}
\newcommand{\pg}[1]{\left\{#1\right\}}

\newenvironment{system}%
{\left\lbrace\begin{array}{@{}l@{}}}%
{\end{array}\right.}
\allowdisplaybreaks

\def\mc{\mathcal}
\def\mb{\mathbb}
\def\rm{\mathrm}

\def\be{\begin{equation}}
	\def\ee{\end{equation}}
\def\ba{\begin{eqnarray}}
	\def\ea{\end{eqnarray}}

\def\p{\partial}

\def\a{\alpha}

\def\g{\gamma}
\def\G{\Gamma}
\def\d{\delta}

\def\l{\lambda}

\def\m{\mu}
\def\n{\nu}
\def\o{\omega}
\def\O{\Omega}
\def\r{\rho}
\def\s{\sigma}
\def\S{\Sigma}

\def\th{\theta}

\def\ra{\rightarrow}

\begin{document}
\maketitle

\noindent ABSTRACT: Understanding quantum dynamics in curved spacetime is a central challenge at the intersection of quantum mechanics and gravity. Anti-de-Sitter (AdS) spacetime plays a pivotal role in the context of the AdS/CFT correspondence, which relates gravitational dynamics in the AdS bulk to a conformal field theory (CFT) living on its boundary. Despite its foundational importance, direct experimental access to dynamical quantum phenomena in Lorentzian AdS spacetime has so far remained out of reach. Here, we report the first experimental emulation of fermionic wave packet dynamics in Lorentzian AdS spacetime using a photonic platform. By mapping the Dirac equation in curved spacetime onto the propagation of light in engineered wave\-guide arrays, we directly observe gravitational confinement of relativistic wave packets and resolve their center-of-mass motion in real time. We identify a characteristic superposition of slow geodesic oscillations governed solely by spacetime curvature and fast Zitterbewegung arising from relativistic particle--antiparticle interference. While the geodesic frequency is independent of fermion mass, the Zitterbewegung frequency exhibits a distinct joint dependence on mass and curvature, revealing a curvature-induced modification of relativistic quantum dynamics. Our results provide the first quantitative experimental access to fermionic bulk dynamics in emulated AdS$_2$ spacetime with Lorentzian signature.  This establishes a scalable analog platform that may potentially be used for exploring dynamical aspects of  holography.\\

\noindent ${}^\ast$ {\small  Both these authors contributed equally to this work, J.~H.~focusing on the experimental and C.~B.~on the theoretical aspects.}

\newpage
\section{Introduction}
Quantum field dynamics in curved spacetime \cite{Birrell1982,Wald1994} lie at the core of gravity and cosmology. 
A central framework in the context of spacetimes with negative curvature is the Anti-de Sitter/conformal field theory (AdS/CFT) correspondence -- the paradigmatic example of holographic duality -- which relates a gravitational theory in a negatively curved hyperbolic AdS spacetime (the ``bulk'') to a strongly interacting CFT living on its lower-dimensional boundary \cite{Maldacena1999,Gubser:1998bc,Witten1998}, for a review see \cite{Ammon_Erdmenger_2015}. This bulk--boundary correspondence provides a powerful conceptual bridge between gravity and quantum matter.

While many aspects of dynamics in
spacetimes with positive curvature have been emulated in experiments (for reviews see for instance \cite{Barcelo:2005fc,Barcelo2011AnalogueGravity,Braunstein2023AnalogueQuantumGravity}), for spacetimes with negative curvature there are significantly less examples. These include emulations of hyperbolic planes with Euclidean signature using quantum circuits 
\cite{Kollar2019HyperbolicLattices, Lenggenhager2022,Zhang2022HyperbolicTopologicalStates,Chen2026SpaceTimeTopologies}
or superconducting wires \cite{Dede:2026sjc}.

A new feature that we introduce in the present paper is an experimental emulation of aspects of dynamics in a 1+1-dimensional AdS space with Lorentzian signature, in which the time direction is emulated by propagation along waveguides.
%\cite{Birrell1982}, yet remain largely inaccessible to direct experimental tests \cite{Wald1994,Barcelo:2005fc}. 
%, but its dynamical bulk predictions have remained almost entirely untested \cite{Aharony2000} \textcolor{blue}{Change}. 
One distinctive consequence of AdS dynamics  is gravitational confinement: timelike geodesics and wave packets do not escape to infinity, but instead periodically refocus toward the center \cite{Calabi1962RelativisticSpaceForms}. For massive particles this implies a striking dynamical signature -- injected wave packets reverse their motion and execute oscillations with a frequency that is determined solely by the spacetime curvature, while the amplitude decreases with increasing mass. We experimentally realize this behavior by emulating the propagation of fermions in 1+1-dimensional AdS space, building on the fermion geodesics discussed in 
 \cite{Tho2016}.
%\textcolor{blue}{Change. This paper talks about geodesics, but not about wavepackets.}

 Beyond geodesic confinement, relativistic fermions exhibit Zitterbewegung arising from coherent interference between particle and antiparticle components of the Dirac field \cite{Schrodinger1930,PhysRevLett.105.143902,Gerritsma2010,Kobakhidze_2016}. 
%In flat spacetime, this contribution can be removed from physical observables by a global unitary transformation associated with a well-defined energy operator and is therefore usually regarded as unphysical \cite{Foldy1950}. In curved spacetime, however, no such global transformation exists in general, making particle--antiparticle mixing an intrinsic dynamical effect\cite{Kobakhidze_2016} of such systems. 
We show theoretically and demonstrate experimentally that AdS spacetime realizes a regime in which this mixing does not only persist in a coherent fashion, but is actually further amplified by the presence of curvature. %\co{While Zitterbewegung can be an observable dynamical effect in non-static spacetimes\cite{Kobakhidze_2016}, it remains to be clarified in the case of Anti-de-Sitter spacetime.}

As our central result, we report the experimental emulation of fermionic wave-packet dynamics in an analog of (1+1)-dimensional Lorentzian AdS spacetime. Using a photonic waveguide platform \cite{Davis1996,Szameit_2010} that directly emulates the Dirac equation in curved spacetime, we access real-time bulk dynamics beyond Euclidean analog implementations \cite{Lenggenhager2022,Dey2024,Chen}, which do not realize covariant relativistic dynamics. In doing so, we observe a coherent superposition of two oscillatory motions: a slow geodesic undulation whose frequency depends exclusively on the AdS curvature and is independent of the fermion mass, and a faster Zitterbewegung whose frequency in turn depends on both mass and curvature (see Fig.\ref{fig:1} for a conceptual sketch). Both scalings quantitatively agree with our gravitational predictions and with the mapping between spacetime parameters and waveguide couplings. Our results establish an experimental route toward probing  dynamical phenomena in Anti-de Sitter space underlying holographic duality, and more generally provide a scalable platform for exploring nonequilibrium dynamics in emulated curved spacetimes.
\section{Results}
\begin{figure}[h!]
\centering
\includegraphics[width=\linewidth]{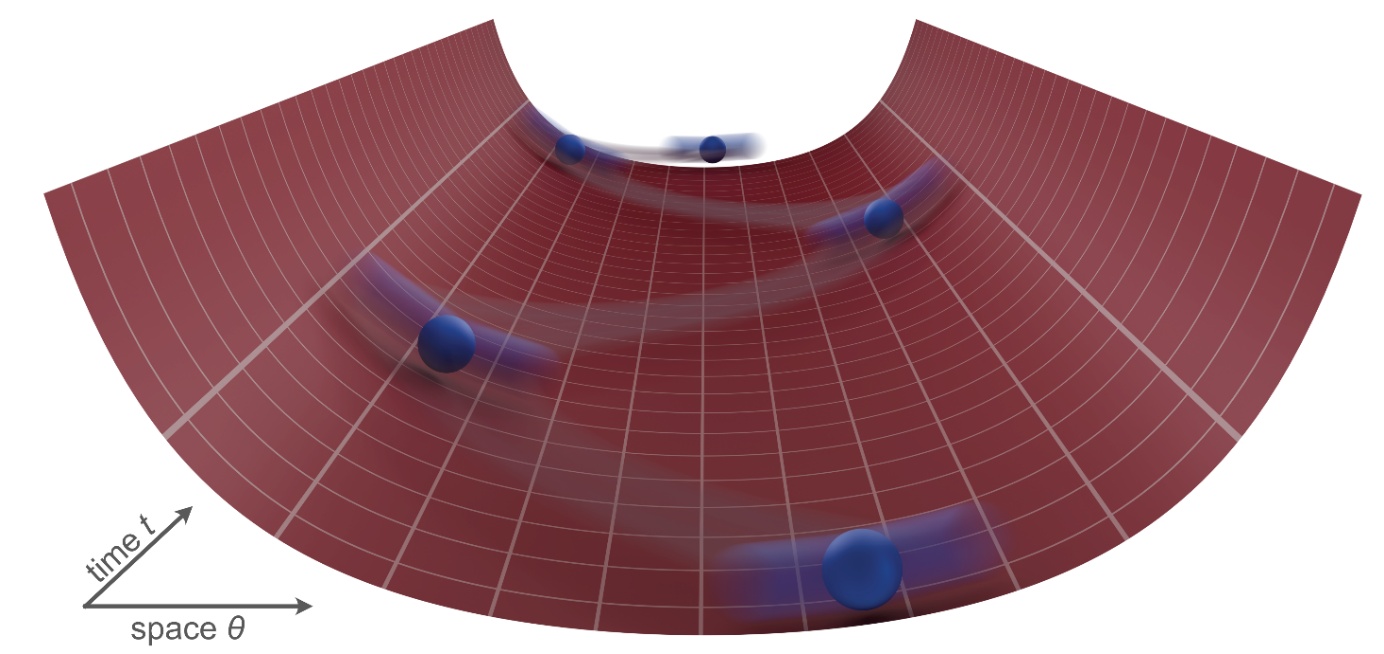}
\caption{A fermionic particle in hyperbolic AdS\(_2\) spacetime. A 2-dimensional coordinate system with a time coordinate \(t\) and a spatial coordinate \(\theta\) is used. The dynamic of a fermion (blue sphere) in AdS spacetime (red surface) is described by the superposition of two distinct oscillations; a slow geodesic motion resulting from the curvature of spacetime and the fast Zitterbewegung, a jittering motion, resulting from the superposition of positive and negative eigenenergies of the Dirac fermion itself.}
\label{fig:1}
\end{figure}

Using the two-dimensional coordinate system \({t, \theta}\), where \(t\) denotes time and the spatial coordinate \(\theta\in\left[-\frac{\pi}{2},\frac{\pi}{2}\right]\) parametrizes the radial direction of two-dimensional anti-de Sitter spacetime (Fig. \ref{fig:1}), the metric of AdS\(_2\) with curvature radius \(l\) and velocity of light \(c\) is

\begin{equation}
ds^2 = \frac{1}{\cos^2\theta}\left(-c^2dt^2 + l^2d\theta^2\right).
\label{eq:ads_metric}
\end{equation}

The dynamics of Dirac fermions in AdS spacetime are governed by the Dirac equation in curved spacetime, which depends explicitly on the fermion mass \(M\) and the underlying geometry \cite{Green_Schwarz_Witten_2012,blanco2022diracfieldmathrmads2representations}. For the AdS\(_2\) metric, the time evolution of a fermionic wave packet \(\psi\left(t,\theta\right)\) is described by (see Appendix)

\begin{equation}
\frac{i}{c}\partial_t\psi(t,\theta)=\left[-\frac{i}{l}\sigma_x\partial_\theta+\frac{Mc}{\hbar\cos\theta}\sigma_z\right]\psi(t,\theta).
\label{eq:dirac_ads}
\end{equation}
where \({\sigma}_{x}\) and \({\sigma}_{z}\) denote the Pauli matrices and \(\hbar\) the reduced Planck's constant. The wave function \(\psi\left(t,\theta\right)\) is a two-component complex spinor of the form \newline $\psi\left(t,\theta\right)=\left\{\psi^{(1)}(t,\theta),\psi^{(2)}(t,\theta)\right\}^{T}$. Uniquely to AdS spacetime, a generic solution of the Dirac Eq.\eqref{eq:dirac_ads} may be decomposed in discrete positive and negative frequency eigenmodes (see section \ref{secapp:2} of the Appendix). This corresponds to particle and antiparticle states, the interference between which gives rise to fermionic Zitterbewegung. 
We find that the center-of-mass motion of a fermionic wave packet is given by the superposition of Zitterbewegung and geodesic motion, resulting in the associated frequencies (see section \ref{secapp:3} and \ref{secapp:4} in the Appendix)
\begin{equation}
\omega_{\rm{Geo}}=\frac{c}{l},\qquad \omega_{\rm{ZB}}=c\left(\frac{2Mc}{\hbar}+\frac{1}{l}\right).
\label{eq:frequencies}
\end{equation}

\begin{figure}[h!]
\centering
\includegraphics[width=\linewidth]{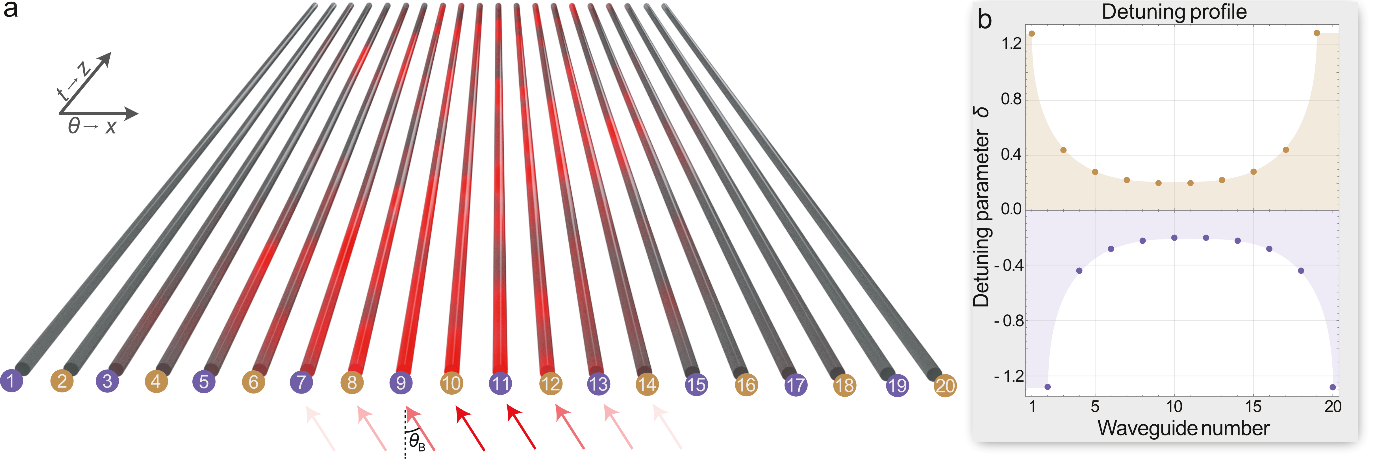}
\caption{Photonic implementation of AdS\(_2\) spacetime dynamics. a The continuous Dirac equation with coordinates \({t,\theta}\) is discretized and mapped to the light evolution in a planar waveguide array with coordinates \(\left\{z,x\right\}\). The constant coupling \(\kappa\) encodes the strength of curvature in AdS\(_2\). To correctly represent Dirac dynamics, a tilted broad excitation under the Bragg angle was used. b The waveguide detuning \(\delta(\theta)\) depends on the position in the array and encodes the geometry of curved space time and follows in the case of AdS\(_2\) a \(1/\cos\left(\theta\right)\) profile, in contrast to an alternating but flat detuning profile in flat space.}
\label{fig:2}
\end{figure}

While both frequencies change with the AdS curvature radius \(l\), only the Zitterbewegung frequency depends also on the fermion mass \(M\). The result \eqref{eq:frequencies} indicates that Zitterbewegung in AdS spacetime indeed manifests as a coherent oscillation. In the flat-space limit \(l\to\infty\), the Zitterbewegung frequency correctly reduces to the well-known \cite{Schrodinger1930,PhysRevLett.105.143902,Gerritsma2010} \(\omega_{\rm{ZB}}^{\rm flat}=2Mc^2/\hbar\).
To emulate Dirac dynamics in AdS\(_2\) using a photonic platform, we map the continuous Dirac equation onto a discrete coupled-mode description of light propagation in a waveguide array. We employ a planar bipartite waveguide lattice consisting of alternating high- and low-index waveguides, as illustrated in Fig. \ref{fig:2}a. The evolution of the electric field amplitude \({a}_{n}\) in waveguide \(n\) along the propagation direction \(z\) is governed by the coupled-mode equations \cite{PhysRevLett.105.143902,zitterLonghi_2010}

\begin{equation}
i\partial_z a_n = \kappa_n\left(a_{n-1}+a_{n+1}\right)+(-1)^n\delta_n a_n,
\label{eq:coupled_mode}
\end{equation}
where \({\kappa}_{n}\) denotes the coupling between adjacent waveguides and \({\delta}_{n}\) represents the on-site detuning. Light propagation in the array is described in the two-dimensional coordinate system \(\{x,z\}\), where the propagation coordinate \(z\) is continuous and the transverse coordinate \(x=dn\) is discrete, with \(d\) as the waveguide spacing and \(n\in \mathbb{Z}\). In the mapping, propagation along \(z\) emulates the time evolution of the Dirac spinor, while the discrete transverse coordinate \(x\) represents the AdS spatial coordinate \(\theta\), such that \(ct\to z\) and \(\theta\to x\) (Fig. \ref{fig:2}a). %\tc{orange}
{In flat spacetime}, a crucial ingredient of the photonic emulation is the dispersion near the edge of the Brillouin zone, which enables an effective Dirac description \cite{PhysRevLett.105.143902}. Consequently, the incident light must be confined to a narrow momentum range around transverse wave vector \(k=\pi\). Experimentally, this condition is fulfilled by a tilted Gaussian excitation injected at the Bragg angle, as shown in Fig. \ref{fig:2}a.
%\tc{orange}
{As the spatial translation along the $\th$ coordinate is broken in AdS$_2$, the notions of dispersion relation and Brillouin zone were generalized to the case of hyperbolic spaces \cite{Maciejko_2021}. Here, the discretization scheme that we describe below provides a real space mapping between the discretized Dirac equation and the coupled mode equations. It also ensures that, in the small curvature limit $ l \gg 1 $, we recover the dispersion relation and the effective Dirac description of the flat space case.}

As detailed in section \ref{secapp:4} of the Appendix, we discretize the AdS coordinate as \({\theta}_{s}=2\pi s/N\), where \(N\) is the total number of waveguides in the array and \(s\in \mathbb{Z}\) is the discretization number \cite{Koke_2016}. Each discretized position \({\theta}_{s}\) is represented by a pair of waveguides, \({a}_{2s+N/2}\) and \({a}_{2s+N/2+1}\), corresponding to the two components of the Dirac spinor. The resulting discretized Dirac equation in AdS\(_2\) assumes the same form as Eq. \eqref{eq:coupled_mode}, such that we find the relations \(\kappa\leftrightarrow N/(2\pi l)\) and \(\delta_n\leftrightarrow Mc/(\hbar\cos\theta_n)\). This mapping establishes a direct correspondence between the photonic lattice parameters and the spacetime geometry: the uniform coupling \(\kappa\) encodes the global curvature of AdS\(_2\), while the spatially varying detuning profile \(1/\cos\left({\theta}_{n}\right)\) reproduces the AdS geometry and its offset \({\delta}_{0}\) sets the fermion mass \(M\). The center-of-mass motion of the solutions of Eq. \eqref{eq:coupled_mode} corresponds to the same superposition of two oscillations, with frequencies identical to those in Eq. \eqref{eq:frequencies} for the continuous case. The resulting detuning landscape implemented in the experiment is shown in Fig. \ref{fig:2}b.
With the discretized Dirac equation in AdS\(_2\) established, we experimentally implement the corresponding photonic lattice using femtosecond laser inscription in fused silica samples \cite{Davis1996,Szameit_2010}. Owing to the strong confinement of the dynamics toward the center of the array, the outermost pairs of waveguides remain unpopulated such that only 16 out of the 20 waveguides contribute to the evolution and are, hence, omitted in the fabrication. The propagation of light in the array is monitored via fluorescence microscopy, which provides a continuous measurement of the light intensity in each waveguide and directly maps the probability distribution of the fermionic wave packet.

\begin{figure}[!h]
\centering
\includegraphics[width=\linewidth]{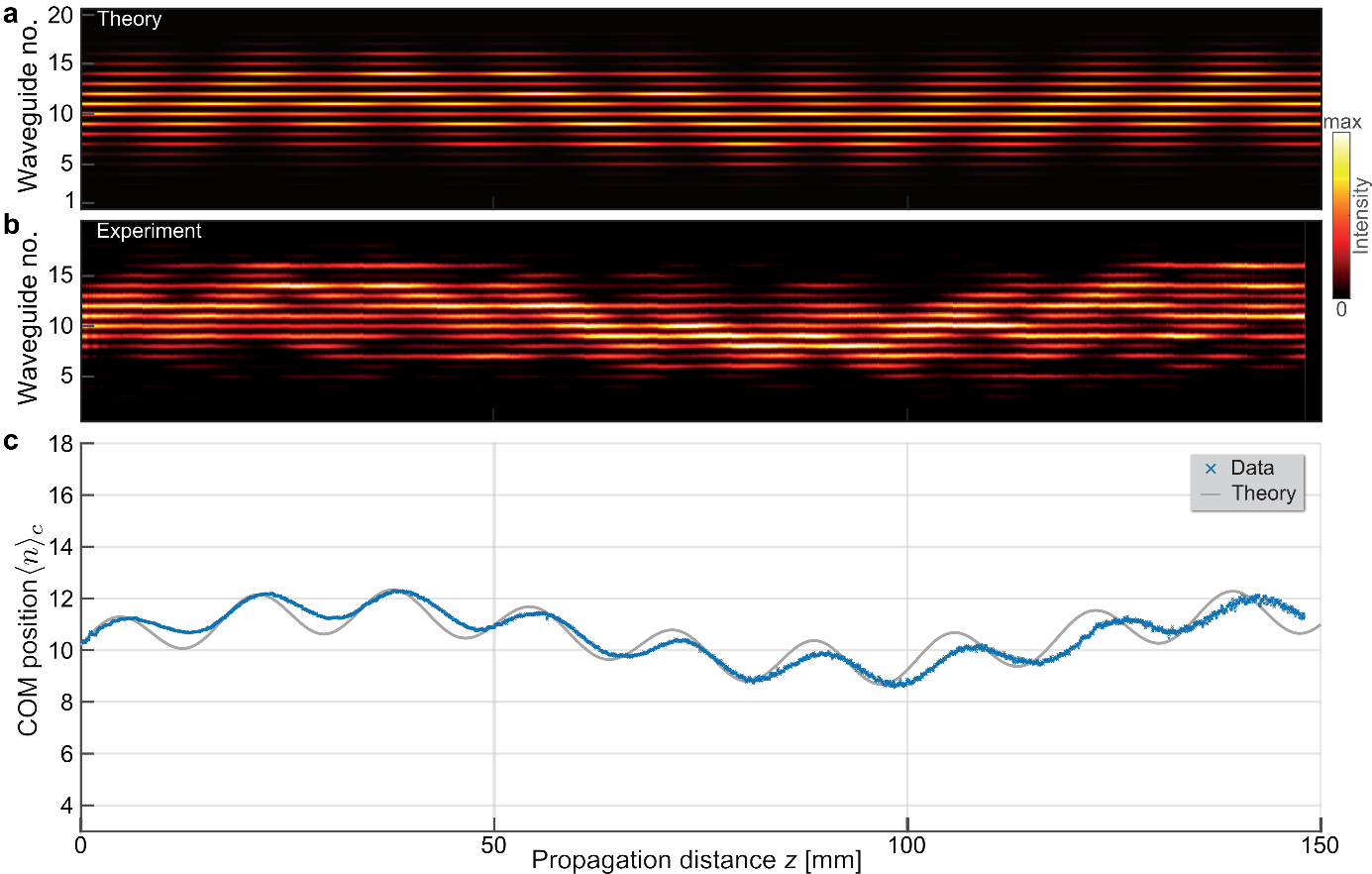}
\caption{Experimental observation of spacetime dynamics. a Simulation of the evolution of a tilted Gaussian excitation in the hyperbolic spacetime array with \(\kappa=0.19\) mm\(^{-1}\) and \({\sigma}_{B}=0.16\) mm\(^{-1}\). b Observed fluorescence image of the evolution of a tilted Gaussian excitation in the fabricated waveguide array with the same parameters, showing a very good overlap with the simulation. c Center-of-mass movement of the light beam clearly showing the two superimposed oscillations with frequencies \({\omega}_{\rm{Geo}}\) and \({\omega}_{\rm{ZB}}\) with a good match to the theory. In our analytical comparison of the COM motion of the Dirac fermion in continuous AdS spacetime \({\left\langle \theta \right\rangle}_{\psi}\) and the light beam in the discretized lattice \({\left\langle n \right\rangle}_{c}\), we also derive a correction term accounting for the minor discrepancy between theory and experiment (see section 4.1 of the Appendix).}
\label{fig:3}
\end{figure}

We realize a waveguide array with a detuning offset of \({\delta}_{0}=0.16\)~mm\(^{-1}\), corresponding to the fermion mass, and a uniform coupling strength of \(\kappa=0.19\)~mm\(^{-1}\), encoding a curvature radius of \(l=16.75\cdot{10}^{2}\) mm. The array is excited by a tilted Gaussian beam of wavelength \(\lambda=633\)~nm, spanning approximately eight waveguides. The Bragg angle is calibrated using a homogeneous lattice, for which such an excitation coincides with minimal diffraction. Figures \ref{fig:3}a,b compare the measured fluorescence signal with numerical simulations of the coupled-mode equations. As predicted by our theory for the emulated fermion mass, the wave packet does not reach the boundary of the array. Instead, we observe the foreseen oscillation about the center of the array due to geodesic confinement in AdS spacetime, superimposed with a higher-frequency oscillatory motion corresponding to Zitterbewegung. The experimental data show excellent agreement with theoretical prediction \eqref{eq:frequencies} and numerical simulations.

\begin{figure}[!h]
\centering
\includegraphics[width=\linewidth]{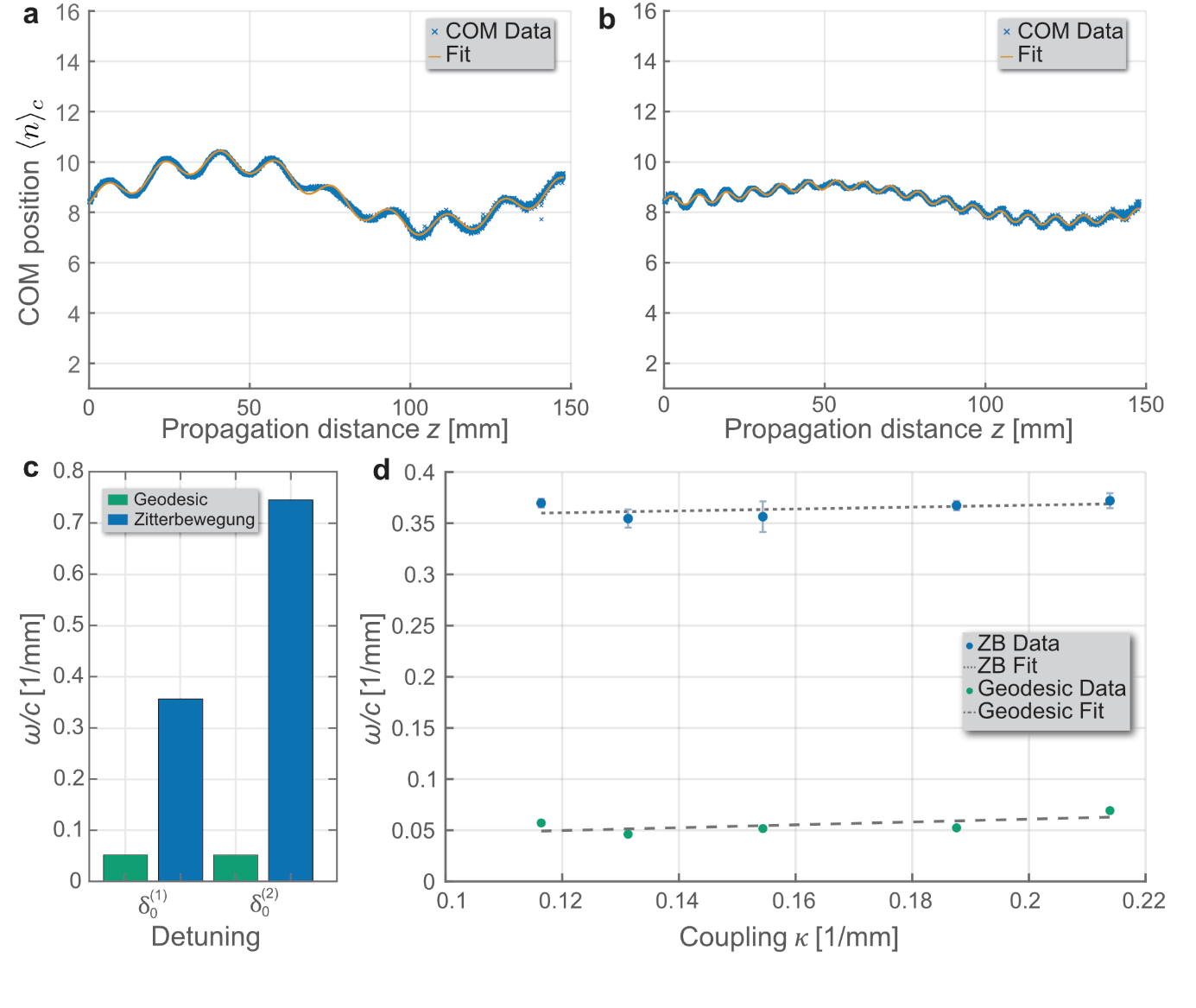}
\caption{Frequency dependence. COM movement for a detuning offset \({\delta}_{0}^{(1)}=0.16\)~mm\(^{-1}\) corresponding to mass \({M}_{1}\) and b \({\delta}_{0}^{(2)}=0.37\)~mm\(^{-1}\) corresponding to mass\({M}_{2}\). While the geodesic frequency stays constant when the mass is changed, the ZB frequency clearly increases with mass, as shown in c. d Frequencies for constant detuning \({\delta}_{0}^{(1)}\) and varying coupling constant. Both frequencies increase with increasing coupling and thus curvature.}
\label{fig:4}
\end{figure}

To quantitatively analyze the fermionic dynamics in AdS\(_2\), we investigate the center-of-mass (COM) motion of the wave packet. In the continuum description, the COM is defined as the expectation value of the spatial coordinate, \(\langle\theta\rangle_{\psi}=\sum d\theta\,\theta|\psi|^2\), which is mapped in the photonic lattice to the COM of the light field, \(\langle n\rangle_c=\sum_n n|a_n|^2\). The measured COM dynamics, shown in Fig. \ref{fig:3}c, clearly reveal a superposition of two components: a fast oscillation at the Zitterbewegung frequency \({\omega}_{ZB}\) and a slower one at the geodesic frequency \({\omega}_{\rm{Geo}}\), in excellent agreement with the theoretical analysis of the coupled-mode equations \eqref{eq:coupled_mode} and the frequency relations \eqref{eq:frequencies}.
\newpage To study the effect of fermion mass at constant curvature, we fabricate two waveguide arrays with detuning offsets \({\delta}_{0}^{(1)}=0.16\)~mm\(^{-1}\) and \({\delta}_{0}^{(2)}=0.37\)~mm\(^{-1}\), corresponding to two different fermion masses \({M}_{1}\) and \({M}_{2}\) , for a fixed coupling strength of \(\kappa\approx 0.17\)~mm\(^{-1}\) (Fig. \ref{fig:4}a,b). As shown in Fig. \ref{fig:4}c, increasing the fermion mass leads to a clear increase of the Zitterbewegung frequency, while the geodesic frequency remains unchanged. The Zitterbewegungs frequency \({\omega}_{\rm{ZB}}\) is obtained from the Fourier transform of the measured COM trajectories, whereas \({\omega}_{\rm{Geo}}\) is extracted by fitting a superposition of slow and fast oscillations to the COM data, providing \({\omega}_{\rm{ZB}}\) as an initial condition.
Finally, to tune the AdS curvature radius at constant mass, we fabricate five additional waveguide arrays with coupling strengths ranging from \(\kappa=0.12\) mm\(^{-1}\) to \(\kappa=0.21\)mm\(^{-1}\) and constant detuning \({\delta}_{0}^{(1)}\). The resulting dependence of both oscillation frequencies on the coupling strength is shown in Fig. \ref{fig:4}d. In agreement with our theoretical predictions, both frequencies increase with increasing curvature, with slopes close to the expected value of \(2\pi/N\).

\newpage In summary, we have experimentally realized fermionic wave-packet dynamics in a Lorentzian analog of two-dimensional anti-de Sitter spacetime and directly observed key dynamical signatures of curved spacetime physics. By mapping the Dirac equation in AdS to a photonic waveguide platform, we accessed real-time bulk dynamics and demonstrated that the center-of-mass motion of a Dirac wave packet is governed by a coherent superposition of two oscillatory contributions: a slow geodesic oscillation induced by gravitational confinement through spacetime curvature and a fast Zitterbewegung arising from particle--antiparticle interference. We quantitatively extracted both frequencies and established their distinct dependence on curvature and fermion mass, thereby providing the first experimental validation of fundamental dynamical predictions of AdS bulk physics. Our results %\tc{orange}
{based on analog methods } establish a controlled experimental testbed for future investigations of dynamical %\tc{orange}
{quantum phenomena} and nonequilibrium dynamics in negatively curved spacetimes.

While the present work focuses on bulk dynamics in AdS\(_2\), an important next step will be to extend this work {to higher-dimensional spacetimes, in particular to AdS$_3$, and establish a mapping between the propagation of fermions in AdS$_3$ and the propagation of light in a two-dimensional waveguide array. This will involve the discretization of  the Dirac equation in AdS$_3$ using the discretization scheme proposed in \cite{Koke_2020} for the case of 2+1-dimensional spacetimes.  %\textcolor{red}
{Moreover, modulating the refractive index along the waveguides will allow for the generation of off-diagonal metric components and hence the emulation of real-time dynamics in Lorentzian AdS$_3$ geometries.} We also plan to establish a similar set-up for studying the  propagation of photons in emulated AdS spacetimes.
} 
%\textcolor{blue}{modulation along waveguides}

Furthermore, in the spirit of the AdS/CFT correspondence where bulk dynamics encode boundary observables, it will be of interest to probe the wave-packet behavior near the AdS boundary and to investigate the role of different boundary conditions. More broadly, our platform enable studies of stability and mass bounds in curved spacetime, and potentially unitarity bounds for fermions in AdS/CFT \cite{Foit2020} and Breitenlohner-Freedman stability bounds for scalars \cite{Basteiro2023}.

One particularly interesting result that we find is {the close agreement between the theoretical prediction and the experimental observation of the Zitterbewegung frequency.  While in general, Zitterbewegung in AdS$_2$ is an infinite superposition of oscillations with frequency labeled by a positive integer $n$, $\o_\pm[n]=\pm \pr{c\pr{\frac{2c M}{\hbar}+\frac{1+2n}{l}}}$ (see section \ref{secapp:3} of the Appendix), experimentally only the first mode is observed. Nevertheless, the observed experimental results already features the expected motion of a fermion in AdS$_2$ quite well, up to a correction discussed in section \ref{app:res} of the Appendix.} We leave a detailed analysis on Zitterbewegung for quantum fields in AdS to the future, as well as its imprint on the dual field theory that may constitute an observable allowing for a test of the AdS/CFT correspondence.

%\tc{orange}{Besides, as the Zitterbewegung affects the dynamics of the fermions in curved spacetimes, it could be interesting to understand more precisely in which case (static-non-static spacetimes,...) it  actually describes an observable effect. It could indeed introduce a previously unexplored ingredient into holographic fermion dynamics and opens the possibility to study how particle--antiparticle interference manifests itself in boundary correlation functions and information transport.}

 Finally, the experimental setup presented, involving an array of coupled waveguides of different refraction index,  may be generalized straightforwardly from single-particle analogs to genuinely quantum states of light, such as indistinguishable or entangled photons. In this way,  the approach presented opens a route toward experimentally exploring quantum correlations, entanglement dynamics and thermalization processes in curved spacetime \cite{Terashima2004,Palmer_2012}. This establishes a bridge between analog gravity and current efforts to experimentally access holographic dynamics, information flow and thermalization in quantum many-body systems.

\section{Sample fabrication and characterization}

We inscribed the photonic wave guide arrays using femtosecond laser-direct writing14. Ultrashort laser pulses of 270 fs duration from a frequency-doubled fiber laser system (Coherent Monaco) at a wavelength of 517 nm and a repetition rate of 333 kHz were focused into a 150 mm \(\times\) 25 mm \(\times\) 1 mm fused silica chip (Corning 7980) by means of a microscope objective (\(\times\)50, numerical aperture = 0.6). The sample was positioned with 50 nm precision by a three-axis motorized translation stage (Aerotech ALS180).

The coupling strength between waveguides was varied, by changing the distance between them, whereas the on-site detuning was tuned by changing the inscription speed of the individual waveguide. Characterization scans beforehand the fabrication of the arrays ensured that the desired coupling and onsite detuning were met. After each of the arrays, one directional coupler with equal coupling distance and writing velocity as used in the middle of the respective array was inscribed. This coupler was used to determine the implemented coupling and detuning offset of the probed array.

The arrays were characterized using coherent light from a 633nm Helium-Neon laser (Melles Griot - 25mW) and the intensity distribution in the waveguides from the top of the sample was measured by fluorescence microscopy14, which is enabled by color centers formed in the inscription process, using a CMOS camera (Basler Ace). The excitation at the edge of the Brillouin zone was ensured using a slit placed in Fourier space in front of the sample. Varying the width of the slit enables control of the excitation width, changing its horizontal position control of the excitation angle. The \(1/{e}^{2}\)-width of the excitation was chosen to excite approximately 8 waveguides. The Bragg angle is calibrated using a homogeneous lattice, for which such an excitation coincides with minimal diffraction.

The fast Zitterbewegungs frequency \({\omega}_{\rm{ZB}}\) is extracted from the measured COM using a Fourier transform of the data and performing a Gaussian fit on the obtained peak. The error of the frequency is obtained by the confidence intervals of the fit parameters. The slow geodesic frequency \({\omega}_{\rm{Geo}}\) is extracted by fitting a superimposed oscillation to the data, providing the priory obtained \({\omega}_{\rm{ZB}}\) as an initial condition.

\section{Data availability}

All data is available upon reasonable request.

\section{Acknowledgements}

AS acknowledges funding from the Deutsche Forschungsgemeinschaft (grants SZ 276/9--2, SZ 276/19--1, SZ 276/20--1, SZ 276/21--1, SZ 276/27--1, and GRK 2676/1--2023 `Imaging of Quantum Systems', project no. 437567992). AS also acknowledges funding from the Krupp von Bohlen and Halbach Foundation as well as from the FET Open Grant EPIQUS (grant no. 899368) within the framework of the European H2020 program for Excellent Science. AS and MH acknowledge funding from the Deutsche Forschungsgemeinschaft via SFB 1477 `Light--Matter Interactions at Interfaces' (project no. 441234705). TAWW is supported by a European Commission Marie Sklodowska-Curie Actions Individual Fellowship (project no. 895254). M.E. acknowledges funding from the German Academy of Natural Sciences Leopoldina (grant number LPDS 2025-02). CB, SH, JE and RM acknowledge financial support by the Deutsche Forschungsgemeinschaft (DFG, German Research Foundation) through the W{\"u}rzburg-Dresden Cluster of Excellence ctd.qmat -- Complexity, Topology and Dynamics in Quantum Matter (EXC 2147, project-id 390858490). R.M. furthermore acknowledges hospitality from the Shanghai Institute
for Mathematics and Interdisciplinary Sciences (SIMIS) and associated travel support under
STCSM Grant 25HB2701900. K.C.M. acknowledges support from the European Commission through a Marie Sklodowska-Curie Actions Individual Fellowship (project No. 101152619). R.N.D. is supported by the PRIME program of the German Academic Exchange Service (DAAD) with funds from the German Ministry of Research, Technology and Space (BMFTR). THL acknowledges funding from the German Federal Ministry of Research, Technology and Space (BMFTR) via the project Qecs (FKZ: 13N16272).
\newpage
\appendix
\section{Appendix:
Theoretical Analysis
}

%\paragraph{}
\noindent Here we discuss the theoretical aspects of the emulation of Anti-de-Sitter spacetime using coupled waveguide arrays. Our main result is that the propagation of a classical beam of light in a one-dimensional waveguide array can emulate the propagation of fermions in a two-dimensional Anti-de-Sitter spacetime, $\mathrm{AdS}_2$. We first introduce Anti-de-Sitter spacetime in 1+1 dimensions and describe explicitly the motion of free particles. This is the so-called \textit{geodesic} motion. We then explain briefly how to describe the motion of Dirac fermions in a curved spacetime, via the Dirac equation and focus on the case of $\mathrm{AdS}_2$. We discretize the continuous Dirac equation in $\mathrm{AdS}_2$ and obtain coupled mode equations that describe the propagation of light in a one-dimensional waveguide array, with a constant coupling parameter $\kappa$ and a $\frac{1}{\cos\th}$ profile for the detuning $\delta_n$. The system of discrete equations simulate a subset of the solutions $\left\{\psi_0\right\}$ to the continuous Dirac equation. We analytically compute the motion of the corresponding center of mass $\langle\th\rangle_{\psi_0}$, which turns out to be a superposition of two oscillations that can be identified with the geodesic motion and the Zitterbewegung, respectively.

	\subsection{Anti-de-Sitter spacetime in 1+1 dimensions $\mathrm{AdS}_2$\label{secapp:1}}
	
	\paragraph{}The main goal of this appendix is to explain how the spatial propagation of light in a one-dimensional waveguide array can emulate the time evolution of Dirac fermions in a two-dimensional Anti-de-Sitter spacetime, $\mathrm{AdS}_2$. More precisely, we will show how to map the continuous Dirac equation in $\mathrm{AdS}_2$ to the discrete coupled mode equations and determine the waveguide parameters $\kappa_n,\d_n$.
    
    \paragraph{}In this section, we start by introducing Anti-de-Sitter spacetime $\mathrm{AdS}_2$ (strictly speaking, we consider the universal cover of $\mathrm{AdS}_2$) and its \textit{geodesics}, i.e. the trajectories of free particles in that space.
	As reviewed in \cite{Ammon_Erdmenger_2015}, $\mathrm{AdS}_2$ is a hyperbolic spacetime with negative curvature. We use the coordinate system $\left\{t,\theta\right\}$ in which the metric takes the form 
	\begin{eqnarray}
		ds^2=\frac{1}{\cos^2\theta}(-c^2dt^2+l^2d\theta^2)\label{eq:1}
	\end{eqnarray} where $-\frac{\pi}{2}\leq\theta\leq\frac{\pi}{2}$ and $l$ is the curvature radius of AdS. 
	
	\paragraph{}The geodesics of both massive and massless particles in $\mathrm{AdS}_2$ are periodic trajectories about the origin, defined as $(t,\th=0)$\cite{Sokoowski2016TheBA,Tho2016}. More precisely:\\
	
	\begin{minipage}{0.4\textwidth}
		\begin{itemize}
			\item massless particles bounce back and forth between the center of AdS, $\theta=0$, and the boundaries, $\theta=\pm\frac{\pi}{2}$. 
			\item massive particles do not reach the boundaries but oscillate between $\theta=0$ and $|\theta|=|\theta_{tp}|\,<\frac{\pi}{2}$. $\theta_{tp}$ will be called in the following the \textit{turning point}.
	\end{itemize}\vspace{1cm}\end{minipage}
	\begin{minipage}{0.7\textwidth}
\centering\includegraphics[width=0.9\textwidth]{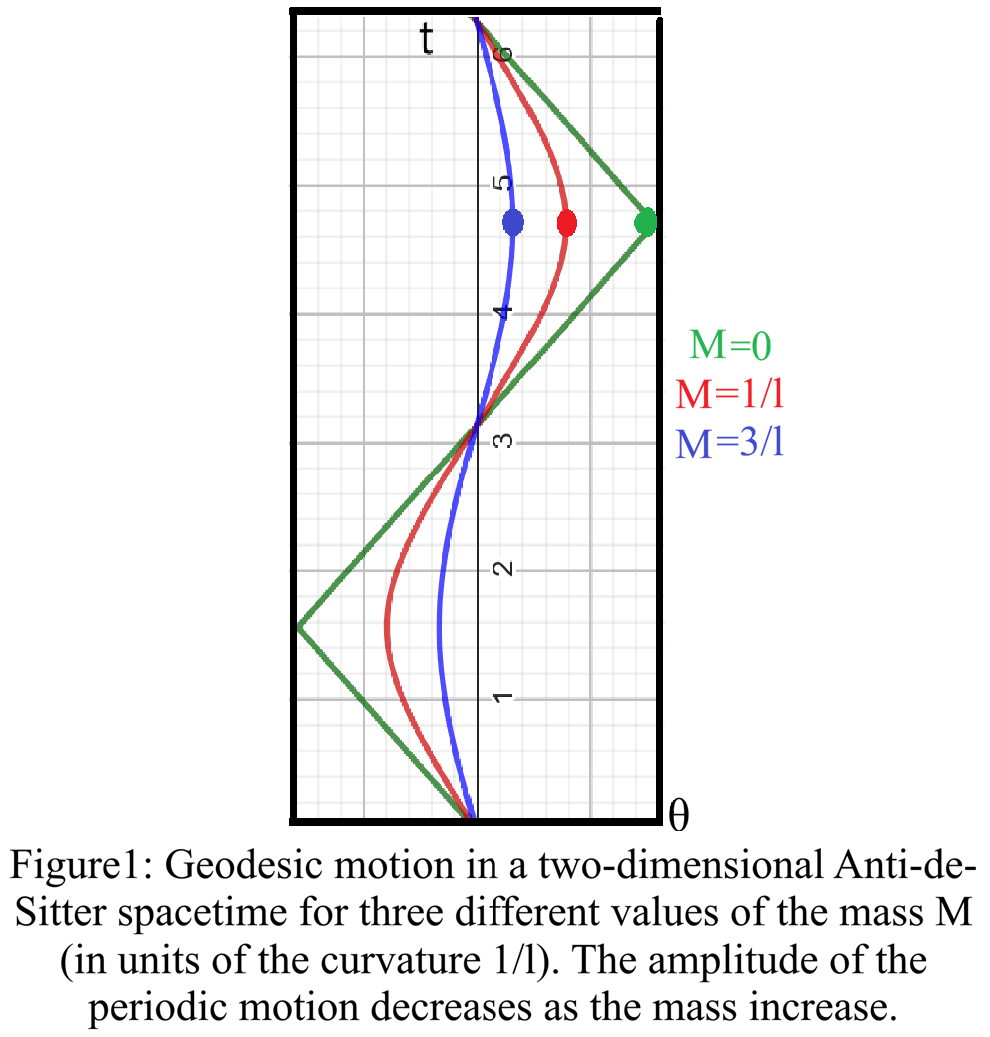}
 
	\end{minipage}
	\paragraph{}We compute the geodesic trajectory associated to the metric (\ref{eq:1}) using the following ingredients \cite{Tho2016}
	
	\begin{itemize}
		\item \underline{Conservation of the four-velocity}: The norm of the four-velocity $u^\m=(\dot{t},\dot{\th})$ is conserved
		\ba u^\m u_\m=\frac{1}{\cos^2\th}(-c^2\dot{t}^2+l^2\dot{\th}^2)=-\delta_{10},\label{eq:36}\ea where $\delta_{01}$ is 0 for massless particles and 1 for massive particles.
		
		\item \underline{Time translation invariance}: The Lagrangian of a free particle of mass $M$ in $\mathrm{AdS}_2$ is given by
		
		\ba \mc{L}=\frac{M}{2}g_{\m\n}\dot{x}^\m\dot{x}^\n=\frac{M}{2\cos^2\theta}(-c^2\dot{t}^2+l^2\dot{\th}^2).\ea
		The Lagrangian does not explicitly depends on time: $\p_t\mc{L}=0$. Consequently, the Euler-Lagrange equations imply that there exists a conserved quantity, $E$
		\ba \p_t\mc{L}&=&\frac{d}{d\l}\frac{\p\mc{L}}{\p\dot{t}}=0,\\
		\Rightarrow E&=&\frac{Mc^2\dot{t}}{\cos^2\th}.\label{eq:35}\ea
		
	\end{itemize} 	
	
	\paragraph{}We solve the equations (\ref{eq:36}) and (\ref{eq:35}) to obtain the geodesic trajectory $t(\th)$
	\ba 
	t(\th)=l \arccot(\frac{\sqrt{1-(\frac{Mc^2}{E})^2-\sin(\theta)^2}}{\sin\th}).
	\ea Those trajectories are shown on the figure above for different values of the masses $M$. 
	
\paragraph{}We invert $t(\theta)$ to obtain the geodesic trajectory $\th_{\mathrm{Geo}}(t)$ of the particle,
\begin{equation}
\th_{\mathrm{Geo}}(t) =\begin{system}
-\arccos\left(\frac{\sqrt{1+(\frac{Mc^2}{E})^2\tan(ct/l)^2}}{\sqrt{1+\tan(ct/l)^2}}\right) \hspace{0.82cm}-\frac{l\pi}{2}\leq ct\leq 0\\
\pi-\arccos\left(-\frac{\sqrt{1+(\frac{Mc^2}{E})^2\tan(ct/l)^2}}{\sqrt{1+\tan(ct/l)^2}}\right) \hspace{0.7cm} 0\leq ct\leq \frac{l\pi}{2}
\end{system}.\label{eq:63}
\end{equation}
This describes a periodic motion with period \begin{eqnarray}
		T=2\pi \frac{l}{c}
	\end{eqnarray} or frequency $\o_{\rm{Geo}}=\frac{c}{l}$ in  SI units. Note that it is independent of the mass $M$ and the energy $E$. Hence, there exists a family of geodesics with the same period but a different amplitude which depends on the rescaled energy $\frac{E}{Mc^2}$.
    
\paragraph{}We compute the Fourier series and find that it is given by a sum of cosines with odd frequencies

\be \th_{\rm{Geo}}(t)=\sum_{\a}c_\a(M,c,E,l) \sin\pr{\frac{2\a+1}{l}t},\,\,\,\a\in\mb N\ ,\ee where the energy dependence is encoded in the Fourier coefficients $c_\a$. Note that $\th_{\rm{Geo}}(t)$ describes the motion of free particles in $\mathrm{AdS}_2$ in the framework of general relativity, which is a classical theory that does not distinguish between fermions and bosons. As we are interested in the propagation of fermions, we will consider the Dirac equation in a $\mathrm{AdS}_2$ background and compute the center of mass motion associated to its solutions. 
	
\subsection{Dirac equation in curved spacetime \label{secapp:2}}

\paragraph{}Let us first recall the Dirac equation in a flat spacetime, namely, how Dirac fermions propagate in spacetimes with vanishing curvature. We follow \cite{blanco2022diracfieldmathrmads2representations,Nakahara:2003nw,Green_Schwarz_Witten_2012} and  set $\hbar=c=1$ (natural units \cite{Thomson_2013}) for simplicity. At the end of the section, we will restore the appropriate factors of $c$ and $\hbar$.
\paragraph{}The Dirac equation reads
	\ba
	\left[\g^a\p_a-M\right]\psi=0,\label{eq:31}
	\ea 
	where $\left\{\g^a\right\}$ satisfy the Clifford algebra
	
	\be 
	\left\{\gamma^a,\g^b\right\}=2\eta_{ab},\label{eq:61}
	\ee and $\eta_{ab}=\mathrm{diag}(-1,1)$ is the Minkowski metric in 1+1 dimensions. 

\paragraph{} A generic curved spacetime is described by a manifold $\mathcal{M}$ equipped with a metric $g_{\m\n}$ and additional matter fields $\Phi$ living on $\mathcal{M}$. At each point $p\in\mc{M}$, the matter field $\Phi(p)$ is living in some vector space $V_p$. Taking derivatives amounts to compare fields at different points, and hence living in different vector spaces $V_p$ and $V_{p+\d p}$. We need to transport the field $\Phi(p)$ living defined at $p$ to the neighboring point, $p+\d p$. This can be done by generalizing the usual partial derivative $\p_\m$ to a \textit{covariant\,derivative} $\nabla_\m$. 

\paragraph{}Let us consider a generic matter field $\Phi(p)$ living in some vector space $V_p$, it can be written as : $\Phi(p)=\Phi^ab_a$ where $\left\{b_a\right\}$ is a basis of $V_p$. The covariant derivative of $\Phi(p)$, $\nabla_\m\Phi(p)$ is given by

\ba 
\nabla_\m\Phi(p)=(\p_\m \Phi^a) b_a+\Phi^a(\O^c_a)_\m  b_c,\label{eq:11}
\ea where
	\begin{itemize}
		\item $\p_\m$ is the usual partial derivative, it gives the variation of the component of the vector $\Phi$.
		\item $(\O^c_a)_\m$ is called a \textit{connection} and implements the transport of the basis $\left\{b_a\right\}$ of $V_p$ to the vector space $V_{p+\d p}$. 
	\end{itemize}
	
\paragraph{}This construction holds generically for any spacetime and hence for flat spacetimes as well. However, as the curvature vanishes in flat spacetimes, there exists a coordinate system such that $\O_\m$ vanishes everywhere.

\paragraph{}In two dimensions, spinors $\psi$ transform in representations $(\r,V)$ of $SO(1,1)$ and not $GL(2,\mathbb R)$. The vector space $V_p$ in which $\psi(p)$ lives should be compatible with the representations of $SO(1,1)$ \cite{Green_Schwarz_Witten_2012}. As a consequence, instead of considering the usual basis $\left\{\p_\m\right\}$ of $T_p M$, the tangent space of $\mc{M}$ at $p$, we perform a change of basis

\be e_a=e_a^\m \p_\m,\ee  where the basis vectors $e_a$ are called \textit{vielbeins}.
These vielbeins satisfy
	\begin{eqnarray}
		\eta_{ab}(p)&=&e_a^\mu(p) e^\nu_b(p)\,g_{\mu\nu}(p),\label{eq:59}
	\end{eqnarray} so that $\mathcal{M}$ looks locally flat in the vicinity of $p$. $SO(1,1)$ acts locally on $\psi(p)$, through local Lorentz transformations.
In this framework, the covariant derivative (\ref{eq:11}) of a spinor $\psi$ is written as

\be \nabla_\m \psi=\p_\m\psi+\frac{1}{2}\o_{\m}^{ab}\S_{ab}\psi,\ee
 where $(\O^c_d)_\m=\frac{1}{2}\o_{\m}^{ab}(\S_{ab})^c_d$ is the connection. $\o_\m^{ab}$ is called the spin-connection and depends on the vielbeins and on the Christoffel symbols as
 \be \o_{\m a}^b=e^b_\n\p_\m e_a^\n +e^b_\n\G_{\m\s}^\n e_a^\s.\label{eq:60}\ee
The matrix $(\S_{ab})^c_d$ is an element of $SO(1,1)$ that acts on the spinor, it is given by \cite{blanco2022diracfieldmathrmads2representations,Green_Schwarz_Witten_2012} 
 \be\S_{ab}= \frac{1}{4}\left[\g^a,\g^b\right].\ee 
	 
\paragraph{}The Dirac equation in a generic curved spacetime can then be written by replacing the usual partial derivative $\p_\m$ by the covariant derivative $\nabla_\m$. This procedure is called \textit{minimal\, coupling}.
\begin{eqnarray}
		\left[\g^\m\nabla_\m-M\right]\,\psi&=&\left[\g^\m\left(\p_\m\psi+\frac{1}{8}\o_{\m}^{ab}\left[\g^a,\g^b\right]\right)-M\right]\,\psi\\&=&\left[e^\m_a \g^a(\p_\m+\frac{1}{4}\o_{\m a b}\g^a\g^b)-M\right]\Psi=0\label{eq:62}
	\end{eqnarray} where the matrices $\g^\m$ are defined as $\g^\m=e^\m_a \g^a$.
To obtain the last line, we have used the commutation relations of $\g$-matrices (\ref{eq:61}). This is the Dirac equation in a generic curved spacetime $\mc{M}$. The choice of the spacetime (and the coordinate system) fixes $\o_\m^{ab}$.

\subsubsection{The Dirac equation in $\mathrm{AdS}_2$}
	
\paragraph{}We now derive the Dirac equation in $\mathrm{AdS}_2$ for a generic choice of $\g^a$ matrices, following \cite{blanco2022diracfieldmathrmads2representations}. We choose the coordinate system that we have introduced in section \ref{secapp:1} but we set for simplicity $c=\hbar=1$. The metric $g_{\m\n}$ then reads
	
	\be  ds^2=\frac{1}{\cos^2\theta}(-dt^2+l^2d\theta^2),\ee
	where $t\in(-\infty,\infty)$ and $-\frac{\pi}{2}\leq\th\leq\frac{\pi}{2}$.
The vielbeins, Christoffel symbols and the spin connection for this metric are easily computed. For instance, for the vielbeins $(e_a)^\m$, we have
	
	\ba 
	e_0=\cos\th\begin{pmatrix}
		1\\0
	\end{pmatrix}\,\,\,\mathrm{and}\,\,\,e_1=\frac{\cos\th}{l}\begin{pmatrix}
		0\\1
	\end{pmatrix}.\label{eq:40}
	\ea 
	We can indeed check that these satisfy the conditions (\ref{eq:59}).
	
    The only non-vanishing components of the spin connection (\ref{eq:60}) are
	\be \o_{10t}=\frac{\tan\th}{l}=-\o_{01t}\label{eq:45}.\ee
	Using the explicit form of the vielbeins (\ref{eq:40}) and the spin connection (\ref{eq:45}), the Dirac equation (\ref{eq:62}) reduces to
	\ba 
	(\g^\m\nabla_\m-M)\Psi&=&\left[\cos\th(\g^0\p_t+\frac{\g^1}{l}\p_\th+\frac{\g^1}{2l}\tan\th)-M\right]\Psi.\label{eq:46}
	\ea
     Performing in addition the rescaling $\Psi=\psi \sqrt{\cos\th}$, we obtain
	\ba
    \left[\gamma^0\partial_t+\frac{1}{l}\gamma^1\partial_\theta\right]\psi(t,\theta)&=&\frac{M}{\cos\theta}\psi(t,\theta).\label{eq:33}
    \ea 
	   This is the Dirac equation in the coordinate system $\left\{t,\th\right\}$ of $\mathrm{AdS}_2$. This equation is valid for any set of $\g$- matrices satisfying the relation (\ref{eq:61}) for the signature\\ $\eta_{\m\n}=\mathrm{diag}(-1,1)$.
\paragraph{} We use the following representation of $\gamma$-matrices $\left\{\g_A^\m\right\}$, 
	\begin{eqnarray}
		&\g_A^0=i\s_z=\begin{pmatrix}
			i&0 \\0&-i
		\end{pmatrix},\,\,\,\g_A^1=-\s_y=-\begin{pmatrix}0&-i\\i&0\end{pmatrix}.
	\end{eqnarray}
	Using this specific choice, the Dirac equation (\ref{eq:33}) then reads
	\ba
	i\p_{t}\psi_A=-\frac{i}{l}\s_{x}\p_{\th}\psi_A+\frac{M}{\cos\th}\s_{z}\psi_A,\label{eq:2}
	\ea
	where $M$ is the mass of the fermion and $l$ is the curvature radius of $\mathrm{AdS}_2$.
Restoring the factors of $c$ and $\hbar$ (going back to the internation system of units SI) this equation (\ref{eq:2}) reads
\be 
i\frac{1}{c}\p_{t}\psi_A=-\frac{i}{l}\s_{x}\p_{\th}\psi_A+\frac{Mc}{\hbar\cos\th}\s_{z}\psi_A.\label{DiracSI}\ee
\subsubsection{Fundamental solutions}
    
\paragraph{}We now study the fundamental solutions of (\ref{DiracSI}).  Those are two-dimensional complex spinors $\psi_A=\begin{pmatrix}
    \psi_A^1\\\psi_A^2
\end{pmatrix}$ that are linear combinations of hypergeometric functions with nice parity properties.
\paragraph{}As shown in \cite{blanco2022diracfieldmathrmads2representations}, it turns out that it is easier to compute the solution to the Dirac equation using a different representation  of $\g$-matrices. The authors of \cite{blanco2022diracfieldmathrmads2representations} have computed explicitly the fundamental solutions $\psi_B$ of the Dirac equation in $\mathrm{AdS}_2$ for the following choice of $\g$-matrices,
	\ba
\g_B^0=i\s_x\,\,\,\mathrm{and}\,\,\,\g_B^1=-\s_z.
	\ea
	For this specific choice, the Dirac equation reads
	\be
	i\frac{1}{c}\p_t\psi_B=-\frac{i}{l}\s_y\p_\th\psi_B+\frac{Mc}{\hbar\cos\th}\s_x\psi_B.\label{eq:21}
	\ee 
	Using the time-translation invariance, we can look for solutions of the form \\$\psi_B=e^{-i\o t}\begin{pmatrix}\Phi_\o^1(\th)\\\Phi_\o^2(\th)\end{pmatrix}$ so that the differential equation can be written as a system of coupled 1st order differential equations. Explicitly, (\ref{eq:21}) becomes
	\ba 
	\frac{\o}{c}\begin{pmatrix}\Phi_\o^1\\\Phi^2_\o\end{pmatrix}=\begin{pmatrix}
		0& \frac{Mc}{\hbar\cos\th}-\frac{1}{l}\p_\th\\
		\frac{Mc}{\hbar\cos\th}-\frac{1}{l}\p_\th&0
	\end{pmatrix}\begin{pmatrix}\Phi_\o^1\\\Phi^2_\o\end{pmatrix}.
	\ea 
We now eliminate one of the components and solve a second-order differential equation for the remaining component. The authors of \cite{blanco2022diracfieldmathrmads2representations} have carried out the whole analysis and obtained the space of normalizable solutions $\left\{\psi_B\right\}$. The form of the solutions depend on the value of the dimensionless parameter $\frac{Mcl}{\hbar}$. For simplicity, we  only consider the case in which $\frac{Mcl}{\hbar}-\frac{1}{2}\notin\mathbb N$ and $\frac{Mcl}{\hbar}>\frac{1}{2}$ in the following.
In this case, the solutions are Gaussian hypergeometric functions (denoted below by $_2F_1$)

\ba 
&\begin{pmatrix}
	\Phi^1_{\o_\pm}\\\Phi^2_{\o_\pm}
\end{pmatrix}=\begin{pmatrix}(2\frac{Mcl}{\hbar}+1) \left(\frac{1-\sin\theta}{1+\sin\theta}\right)^\frac{Mcl}{2\hbar} {}_2F_1\left[l\frac{\omega_{\pm}[n]}{c},l\frac{\omega_{\pm}[n]}{c},\frac{1}{2}+\frac{Mcl}{\hbar},\frac{1-\sin\theta}{2}\right]
\\l\frac{\omega_{\pm}[n]}{c}\,\cos\theta\, \left(\frac{1-\sin\theta}{1+\sin\theta}\right)^\frac{Mcl}{2\hbar} {}_2F_1\left[1+l\frac{\omega_{\pm}[n]}{c},1+l\frac{\omega_{\pm}[n]}{c},\frac{3}{2}+\frac{Mcl}{\hbar},\frac{1-\sin\theta}{2}\right]\end{pmatrix}\label{eq:22}.\hspace{1cm}\,\,\,
\ea 
	%\paragraph{}
The solutions are labeled by a frequency $\o_{\pm}[n]$ which depend on a positive integer $n\in \mathbb N$,
	\ba 
	 \o_{\pm}[n]=\pm c\pr{\frac{1}{2l}+\frac{cM}{\hbar}+\frac{n}{l}},\,\,\,n\in\mathbb N.\label{freq}
	\ea
%\paragraph{}
We now use the following property: whenever two choices of $\g$ matrices, $\left\{\g_A\right\}$ and  $\left\{\g_B\right\}$, are related by a similarity transform $S$ \begin{eqnarray}
	\gamma^\mu_B=S\,\gamma_A^\mu\,S^{-1},\label{eq:32}
\end{eqnarray} 
the corresponding solutions of the Dirac equation (\ref{eq:33}) are related by
	\begin{eqnarray}
		\psi_A&=&S^{-1}\psi_B.\label{eq:4}
	\end{eqnarray}
Our choice $\left\{\gamma_A^\m\right\}$ is related to the choice $\left\{\gamma_B^\m\right\}$ of \cite{blanco2022diracfieldmathrmads2representations} by the similarity matrix
		\ba
		S=\begin{pmatrix}
			\frac{1}{2}&-\frac{i}{2}\\\frac{1}{2}&\frac{i}{2}
		\end{pmatrix}\,\,\,&,&\,\,\,S^{-1}=\begin{pmatrix}
			1&1\\i&-i
		\end{pmatrix}.\label{eq:23}
		\ea
		 Hence, we find the solutions to the Dirac equation (\ref{eq:2}) (for $\frac{Mcl}{\hbar}>\frac{1}{2}$ and $\frac{Mcl}{\hbar}-\frac{1}{2}\notin \mathbb N$) by applying the transformation (\ref{eq:4}), with S given by (\ref{eq:23}), on the solutions $\psi_B$ (\ref{eq:22}). They are two-dimensional spinors
		\ba\psi_A=S^{-1}\psi_B=e^{-i\o_{\pm} t}\begin{pmatrix}
			\phi^1_{\o_{\pm}}\\\phi^2_{\o_{\pm}}
		\end{pmatrix},\ea 
		with components:   
		
		\begin{eqnarray}
			\phi_{\omega_\pm}^1&=&
			(2\frac{Mcl}{\hbar}+1) \left(\frac{1-\sin\theta}{1+\sin\theta}\right)^\frac{Mcl}{2\hbar} {}_2F_1\left[l\frac{\omega_{\pm}[n]}{c},l\frac{\omega_{\pm}[n]}{c},\frac{1}{2}+\frac{Mcl}{\hbar},\frac{1-\sin\theta}{2}\right]\label{eq:64}\\&\,\,&\,+l\frac{\omega_{\pm}[n]}{c}\,\cos\theta\, \left(\frac{1-\sin\theta}{1+\sin\theta}\right)^\frac{Mcl}{2\hbar} {}_2F_1\left[1+l\frac{\omega_{\pm}[n]}{c},1+l\frac{\omega_{\pm}[n]}{c},\frac{3}{2}+\frac{Mcl}{\hbar},\frac{1-\sin\theta}{2}\right],\nonumber\\\phi_{\omega_{\pm}}^2&=&i(2\frac{Mcl}{\hbar}+1) \left(\frac{1-\sin\theta}{1+\sin\theta}\right)^\frac{Mcl}{2\hbar} {}_2F_1\left[l\frac{\omega_{\pm}[n]}{c},l\frac{\omega_{\pm}[n]}{c},\frac{1}{2}+\frac{Mcl}{\hbar},\frac{1-\sin\theta}{2}\right]\label{eq:24}\\&\,\,&-il\frac{\omega_{\pm}[n]}{c}\,\cos\theta \left(\frac{1-\sin\theta}{1+\sin\theta}\right)^\frac{Mcl}{2\hbar} {}_2F_1\left[1+l\frac{\omega_{\pm}[n]}{c},1+l\frac{\omega_{\pm}[n]}{c},\frac{3}{2}+\frac{Mcl}{\hbar},\frac{1-\sin\theta}{2}\right].\nonumber
		\end{eqnarray} 
        We then find  that a generic solution of (\ref{eq:2}) is then given by a linear combination of the form\begin{eqnarray}
			\psi_A(t,\theta)=\sum_{n}e^{-i\omega_{\pm}[n]t}\,b_{\o_{\pm}}\begin{pmatrix}
				\phi^1_{\omega_\pm}(\theta)\\\phi^2_{\omega_\pm}(\theta)
			\end{pmatrix},\label{eq:7}
		\end{eqnarray}
	with frequency (\ref{freq}) \be \o_{\pm}[n]=\pm c\pr{\frac{1}{2l}+\frac{cM}{\hbar}+\frac{n}{l}}.\ee

\subsubsection{Properties of the solutions \label{subsec2.3}}
\paragraph{}We now aim at computing the center of mass motion of the spinor $\psi$ (\ref{eq:7}), $\langle\th\rangle_\psi$. The solutions $\psi_A$ satisfy a number of properties that will considerably simplify the general form of the center of mass motion $\langle\th\rangle_{\psi_A}$. We list them below.

\paragraph{}We first notice that the components $\phi_{\o[n]}^1$ (\ref{eq:64}) and $\phi_{\o[n]}^2$ (\ref{eq:24}) are real and purely imaginary, respectively, 
\ba 
\phi_{\o[n]}^{1*}=\phi_{\o[n]}^1,\\
\phi_{\o[n]}^{2}=i\underbrace{\tilde{\phi}_{\o[n]}^2}_{\in\mathbb R}.
\ea   
%\paragraph{}
In addition, the components $\phi^1_{\o_{\pm}[n]},\,\phi^2_{\o_{\pm}[n]}$ are odd or even under parity, depending on the frequency $\o_{\pm}[n]$. We find that the components with positive frequencies $\o_+[n]$, $\phi^1_{\o_{+}},\,\phi^2_{\o_{+}}$  satisfy
		\begin{itemize}
			\item \underline{$\o_+$(2p)} : \begin{eqnarray}
				& \phi^{(1)}_{\o_+(2p)}(\theta)=\phi^{(1)}_{\o_+(2p)}(-\theta)\label{eq:25},\\
				&\phi^{(2)}_{\o_+(2p)}(\theta)=-\phi^{(2)}_{\o_+(2p)}(-\theta),\label{eq:26}
			\end{eqnarray}
			\item \underline{$\o_+$(2p+1)} : \begin{eqnarray}
				& \phi^{(1)}_{\o_+(2p+1)}(\theta)=-\phi^{(1)}_{\o_+(2p+1)}(-\theta),\label{eq:27}\\
				&\phi^{(2)}_{\o_+(2p+1)}(\theta)=\phi^{(2)}_{\o_+(2p+1)}(-\theta),\label{eq:28}
			\end{eqnarray}
		\end{itemize}
		while the components with negative frequencies $\o_-[n]$ satisfy
		\begin{itemize}
			\item \underline{$\o_-$(2p)} : \begin{eqnarray}
				& \phi^{(1)}_{\o_-(2p)}(\theta)=-\phi^{(1)}_{\o_-(2p)}(-\theta),\label{eq:29}\\
				&\phi^{(2)}_{\o_-(2p)}(\theta)=\phi^{(2)}_{\o_-(2p)}(-\theta),\label{eq:30}
			\end{eqnarray}
			\item \underline{$\o_-$[2p+1]} : \begin{eqnarray}
				& \phi^{(1)}_{\o_-(2p+1)}(\theta)=\phi^{(1)}_{\o_-(2p+1)}(-\theta),\label{eq:34}\\
				&\phi^{(2)}_{\o_-(2p+1)}(\theta)=-\phi^{(2)}_{\o_-(2p+1)}(-\theta).\label{eq:39}
			\end{eqnarray}
		\end{itemize}
		
	\subsection{Center of mass motion of a Dirac fermion \label{secapp:3}}
		
		\paragraph{}In this section, we compute the expectation value $\langle\theta\rangle_{\psi}$ of a generic solution of the form (\ref{eq:7}). It represents the center of mass motion of the Dirac fermion in $\mathrm{AdS}_2$ and exhibits some features of the geodesic motion predicted by General Relativity (see section \ref{secapp:1}).
		
		\paragraph{}A similar computation has already been carried out in flat spacetime in \cite{PhysRevLett.105.143902,zitterLonghi_2010}. They found that the center of mass motion of the Dirac fermion in flat spacetime is given by a straight line with constant slope, that is,  a constant velocity motion. This corresponds to the geodesic motion of a particle in flat space. When the spinor is a superposition of positive and negative frequency solutions, there is an additional higher frequency oscillation on top of the geodesic motion. This oscillation is called \textit{Zitterbewegung} and its frequency is given by $\o_{\mathrm{ZB}}=2M \frac{c^2}{\hbar}$ in the SI units \cite{PhysRevLett.105.143902}.
        
		\paragraph{}Other works were carried out in the case of curved spacetimes in which either the conserved current of the Dirac spinor  $j^\m=\bar\psi \g^\m\psi$ or, in the Hamiltonian formalism, the velocity operator is computed \cite{ZBAdS,Kobakhidze_2016}. In this paper, we compute the center of mass of the Dirac fermion, extending the work of \cite{PhysRevLett.105.143902} to the case of $\mathrm{AdS}_2$. 
	
		\paragraph{}Following \cite{PhysRevLett.105.143902,zitterLonghi_2010}, the center of mass motion of a generic solution $\langle\th\rangle_\psi$ of the form (\ref{eq:7}) is computed as
		\begin{eqnarray}
			\langle\theta\rangle_{\psi}(t)&=&\int_{-\frac{\pi}{2}}^{\frac{\pi}{2}} d\theta\,\theta\,|\psi(t,\theta)|^2\label{eq:41}\\&=&\sum_{\omega[n_1],\omega[n_2]}b_{\omega[n_1]}^*b_{\omega[n_2]} e^{i(\omega[n_1]-\omega[n_2])\,t}\underbrace{\int_{-\frac{\pi}{2}}^{\frac{\pi}{2}} d\theta \,\theta\,(\phi^{*,(1)}_{\omega[n_1]}\phi^{(1)}_{\omega[n_2]}+\phi^{*,(2)}_{\omega[n_1]}\phi^{(2)}_{\omega[n_2]})(\theta)}_{=\alpha(\omega[n_1],\omega[n_2],M)\,\in\,\mathbb R}\nonumber\\&=&\sum_{\omega[n_1],\omega[n_2]}b_{\omega[n_1]}^*b_{\omega[n_2]} e^{i(\omega[n_1]-\omega[n_2])\,t}\alpha(\omega[n_1],\omega[n_2],M)\ ,\label{eq:54}
		\end{eqnarray}where $\omega[n]$ is a notation for $\omega_{\pm}[n]$.

\paragraph{}We now use the properties of the components $\phi^{(1)}_{\omega[n]},\,\phi^{(2)}_{\omega[n]}$ presented in the subsection \ref{subsec2.3} to simplify the sum \eqref{eq:54}. First, thanks to the fact that $\phi^{(1)}_{\omega[n]}$ is real and $\phi^{(2)}_{\omega[n]}$ imaginary, the coefficient $\a(\omega[n_1],\omega[n_2],M)$ is real and symmetric under the exchange $\o[n_1]\leftrightarrow\o[n_2]$
 
 \be
 \alpha(\omega[n_1],\omega[n_2],M)=\alpha(\omega[n_2],\omega[n_1],M).
 \ee

\paragraph{}We then rewrite the center of mass motion as
\begin{eqnarray}
	\langle\theta\rangle_{\psi}(t)&=&2\underset{n_1<n_2}{\sum_{(\omega[n_1],\omega[n_2])}}\Re( b_{\omega[n_1]}^*b_{\omega[n_2]} \,e^{i(\omega[n_1]-\omega[n_2])\,t})\,\alpha(\omega[n_1],\omega[n_2],M)\label{eq:72}\\&\,&+\sum_{\omega[n]}|b_{\omega[n]}|^2 \,\alpha(\omega[n],\omega[n],M).\nonumber
\end{eqnarray}	$\Re(\cdot)$ denotes the real part of the expression in parenthesis.

\paragraph{}Thanks to the parity properties (\ref{eq:25}-\ref{eq:39}), $\a(\omega(n_1),\o(n_2),M)$ vanishes in the following cases,
		
		\begin{eqnarray}
			\a(\o_+(2p),\o_+(2q),M)=0= \a(\o_+(2p+1),\o_+(2q+1),M),\label{eq:44}\end{eqnarray}\begin{eqnarray}
			\a(\o_-(2p),\o_-(2q),M)=0= \a(\o_-(2p+1),\o_-(2q+1),M),\label{eq:47}\end{eqnarray}\begin{eqnarray}
			\a(\o_+(2p),\o_-(2q+1),M)=0= \a(\o_+(2p+1),\o_-(2q),M).\label{eq:48}
		\end{eqnarray}
\paragraph{}Consequently, the second term in $\langle\theta\rangle_{\psi}(t)$ (\ref{eq:72}) vanishes and the center of mass reduces to
\begin{eqnarray}
	\langle\theta\rangle_{\psi}(t)&=&2\underset{n_1<n_2}{\sum_{(\omega[n_1],\omega[n_2])}}\Re( b_{\omega[n_1]}^*b_{\omega[n_2]} \,e^{i(\omega[n_1]-\omega[n_2])\,t})\,\alpha(\omega[n_1],\omega[n_2],M)\ ,\label{eq:3}\\
    &=& 2\underset{n_1<n_2}{\sum_{(\omega[n_1],\omega[n_2])}}\Re(  b_{\omega[n_1]}^*b_{\omega[n_2]} \,e^{i(\omega[n_1]-\omega[n_2])\,t})\,\alpha(\omega[n_1],\omega[n_2],M)
\end{eqnarray}
where the sum only involves the following pairs of frequencies
\begin{itemize}
	\item $\o_{\pm}(2p+1)-\o_{\pm}(2q)=\pm c\pr{\frac{1+2(p-q)}{l}}$ : this contribution comes from the superposition of solutions with frequencies of the same sign and it will give rise to the geodesic motion. Indeed, if we choose for simplicity the coefficients associated to odd frequencies to be purely imaginary, $b_{\o_\pm(2p+1)}\in i\mb R$, and the coefficients associated to even frequencies to be real, $b_{\o_\pm(2p)}\in \mb R$, we get a sum of sine functions with odd frequencies
    \ba \langle\th\rangle_{\psi,\,\rm{Geo}}(t)=\sum_{\a\geq 0} c_\a \sin\left(c\frac{(2\a+1)}{l}t\right)\ .\ea 
    In principle, when the $\pg{c_\a}$ are chosen to be the Fourier coefficients of the geodesic trajectory $\th_{\mathrm{Geo}}$(t) (\ref{eq:63}), this contribution of the center of mass matches with the geodesic motion expected in GR. In particular, the families of geodesics labeled by the energy $E$ of section \ref{secapp:1} can a piori be observed by choosing the $\pg{c_\a}$ to be the corresponding Fourier coefficients. 
	\item $\o_{\pm}(2p)-\o_{\mp}(2q)=\pm c\pr{2\frac{cM}{\hbar}+\frac{1+2(p+q)}{l}}$: this contribution arises from the superposition of solutions with frequencies of opposite sign. This contribution will give rise to the Zitterbewegung. It depends on the mass of the fermion $M$ and on the curvature radius of $\mathrm{AdS}$, $l$. Assuming again $b_{\o_\pm(2p+1)}\in i\mb R$ and $b_{\o_\pm(2p)}\in \mb R$, we obtain
    
    \ba \langle\th\rangle_{\psi,\,\rm{ZB}}(t)&=&\sum_{\a\geq 0} c_\a \cos\left((2\frac{Mc^2}{\hbar}+c\frac{(2\a+1)}{l}t\right)\ .\ea
    
    We recover the flat space result $\omega_{\mathrm{ZB,\,flat}}=2\frac{c^2M}{\hbar}$ \cite{PhysRevLett.105.143902,zitterLonghi_2010} in the flat space limit.
\end{itemize} 
In the experiments, as we will see in the next section, only the lowest frequency modes are observed. Nevertheless, these already capture both the geodesic and Zitterbewegung motion of a Dirac fermion. 
\subsection{Discretization procedure \label{secapp:4}}

\paragraph{}Recall that we want to map the propagation of light in a 1d waveguide array to the propagation of fermions in $\mathrm{AdS}_2$. Practically, this means that we want to map the Dirac equation in $\mathrm{AdS}_2$ (\ref{DiracSI}) to the coupled mode equations that govern the propagation of light in a one-dimensional waveguide array. The latter are given by
\ba 
	i\p_z a_n=\kappa_n(a_{n-1}+a_{n+1})+(-1)^n\delta_n a_n.
\ea
		
The light propagates in a two-dimensional space with coordinates $(x,z)$ where $z$ is continuous and $x=d\,n$ is discrete ($d$ is the distance between two waveguides). The propagation of the light in the $z$ direction (along the waveguide) emulates the time evolution of the spinor while the discrete coordinate x emulates the spatial coordinate $\th$ of $\mathrm{AdS}_2$. In other words, we consider the mapping

\ba 
ct&\rightarrow& z,\\
\th&\rightarrow& x=nd.
\ea

\paragraph{}For this mapping to work, we need to find a discretization scheme that maps the continuous Dirac equation (\ref{DiracSI}) to the coupled mode equations. In this section, we review the discretization scheme introduced in \cite{Koke_2016} and give the correspondence between the waveguide parameters, the mass of the fermion and the curvature of AdS.

\paragraph{}The AdS coordinates $\theta$ is discretized as
	\begin{eqnarray}
		\theta_s=s\,\tilde{d}\,\,\,\mathrm{with}\,\,\,\tilde{d}=\frac{\pi}{N/2}\label{th},
	\end{eqnarray}
where $N$ is the number of waveguides and $s\in\pq{-(\frac{N}{4}-\frac{1}{2}),(\frac{N}{4}-\frac{1}{2})}$ is a half-integer and $\theta_s\in(-\frac{\pi}{2},\frac{\pi}{2})$. This forms a 1 dimensional lattice with discretization step $\tilde{d}$.

\paragraph{}The spinor $\psi(t,\theta)=\begin{pmatrix}
			\psi_1\\\psi_2
		\end{pmatrix}$ is a two dimensional object and we associate one waveguide to each component. Hence, each discretized coordinate $\theta_s$ is associated to two waveguides. The waveguide array is a 1d lattice with unit cells that consist of two waveguides. The coordinate of the unit cell is denoted by $s$ and is in one-to-one correspondence with the spatial coordinate of AdS, $\theta_s$ (\ref{th}). The coordinates of the pair of waveguides in the $s^{\mathrm{th}}$ unit cell are $2s-1$ and $2s$ and they are associated to the same angle $\th_s$.
		
		\paragraph{}More precisely, the spinor is discretized as \cite{Koke_2016}\begin{eqnarray}
			\psi=\begin{pmatrix}
				\psi_1(\theta)\\\psi_2(\theta)
			\end{pmatrix}\rightarrow \begin{pmatrix}
				\psi_1(s)\\\psi_2(s)
			\end{pmatrix}\rightarrow\,\begin{pmatrix}
				a_{2s}=(-1)^s\, \psi_1(s)\\ a_{2s-1}=-i(-1)^s \,\psi_2(s)
			\end{pmatrix}\label{eq:5}.
		\end{eqnarray}
        
		\paragraph{}In the flat space limit, the spatial translation symmetry is restored and we can compute the dispersion relations of the coupled mode equations and the Dirac equation by Fourier transform. This specific Ansatz (\ref{eq:5}) ensures that both dispersion relations match in the flat space limit.  
		
		\paragraph{}Performing the discretization (\ref{eq:5}), we get the discretized Dirac equation in $\mathrm{AdS}_2$ 
\ba i\p_z a_n=i\frac{1}{c}\partial_{t} a_n=-\frac{1}{\tilde{d}\,l}(a_{n+1}+a_{n-1})+(-1)^n\,\frac{Mc}{\hbar\cos\theta_n}\,a_n .\label{CmeSI}\ea
		Those are couple mode equations describing the propagation of light in a 1d waveguide array with the specific waveguide parameters ($\kappa$, $\delta_n$) 
        \begin{eqnarray}
			\kappa\leftrightarrow\,\frac{1}{l\,\tilde{d}}=\frac{N}{2\pi\,l},\,\,\,\,\,\delta_n\leftrightarrow\frac{ Mc}{\hbar\cos\theta_n}.
		\end{eqnarray}
\paragraph{}The coupling parameter $\kappa$ encodes the \say{global} curvature of $\mathrm{AdS}_2$, while $\d_n$ encodes the geometry of AdS (through the $\frac{1}{\cos\th}$ profile) and the mass of the Dirac fermion.

\subsubsection{Comparison between the center of masses of the Dirac fermion $\langle\th\rangle_{\psi}$ and the lightbeam $\langle n\rangle$}  
\label{app:res}

\paragraph{}In order to check that these coupled mode equations actually emulate the propagation of fermions in $\mathrm{AdS}_2$ we show that the center of mass motion of the fermion predicted by the continuous Dirac equation $\langle\th\rangle_{\psi}$ is directly related to the center of mass of the light beam in the array $\langle n\rangle$.

\paragraph{}The center of mass of the light beam is defined as
\ba \langle n\rangle&=&\sum_{n=1}^{N} n\,|a_{n}|^2,\ea
where  $N$ is the number of waveguide, which is  an even number.
   Rearranging this sum in terms of the coordinate of the unit cell $s$, i.e. the discretized spatial AdS coordinate, we get
	\ba
\langle n\rangle&=&\sum_{s=1}^{\frac{N}{2}} (2s-1)|a_{2s-1}|^2+\sum_{s=1}^{\frac{N}{2}} 2s\,|a_{2s}|^2\\&=&2\sum_{s=1}^{\frac{N}{2}}s\pr{|a_{2s-1}|^2+|a_{2s}|^2}-\sum_{s}|a_{2s-1}|^2.
	\ea

Inserting the expressions of $\th_s$ (\ref{th}), $\psi_1(s)$ and $\psi_2(s)$ (\ref{eq:5}), we express $\langle n\rangle$ in terms of the discretized spinor and the discretized AdS coordinate and take the continuum limit. We get

\ba \langle n\rangle_\psi&=&2\sum_{s=1}^{\frac{N}{2}}\frac{\th_s\,N}{2\pi}\pr{|\psi_2(s)|^2+|\psi_1(s)|^2}-\sum_{s}|\psi_2(s)|^2\\&\ra&2\frac{N}{2\pi}\int \frac{d\th}{\frac{2\pi}{N}}\,\th\pr{|\psi_2(s)|^2+|\psi_1(s)|^2}-\int \frac{d\th}{\frac{2\pi}{N}}\,|\psi_2(s)|^2\\&=&2\pr{\frac{N}{2\pi}}^2 \langle\th\rangle_\psi-\frac{N}{2\pi}\int d\th |\psi^2(\th)|^2\label{threln}.\ea

\paragraph{}We see that the center of mass of the light beam is directly linked to the center of mass of the Dirac fermion up to a correction term $\frac{N}{2\pi}\int d\th |\psi^2(\th)|^2$. Note that in the limit of infinitely many waveguides, $N\ra \infty$, this correction is irrelevant since it is subleading in $N$. It appears only when  a finite number of waveguides is considered. We denote by $\langle n\rangle_\psi$ the center of mass of the lightbeam calculated from the discretized spinor $\psi$ thanks to the formula (\ref{threln}).

\paragraph{}We next determine which Dirac spinors are actually emulated by the experimental platform. To this end, we solve numerically the coupled mode equations (\ref{CmeSI}) for an initial discrete Gaussian state and simulate the associated center of mass motion, denoted by $\langle n\rangle_s$. We find that $\langle n\rangle_s$ is a superposition of two oscillations. We then perform a discrete Fourier transform to extract the following frequencies
		\ba 
		\o_1^{(s)}&=&\frac{c}{l},\\
		\o_2^{(s)}&=&\frac{c}{l}+2M\frac{c^2}{\hbar}.
		\ea 
		Schematically, the numerical simulations predict that the center of mass of the lightbeam looks like
		\ba 
		\langle n\rangle_{\mathrm{s}}=a\cos(\o_1^{(s)}t)+b\cos(\o_2^{(s)}t)\label{sim}.
		\ea 
		
\paragraph{}We compare this prediction with the center of mass of a generic Dirac fermion (\ref{eq:3}) and we deduce that the coupled mode equations with a Gaussian initial state only emulate the following Dirac spinors
\be\psi_0=b\,\phi_{\o_-[0]}(\theta)\,e^{-i\omega_-[0] t}+d\,\phi_{\omega_{\pm}[1]}e^{-i\omega_\pm[1] t}+f\,\phi_{\omega_+[0]}e^{-i\omega_+[0] t}\label{sol} \ee 
Those correspond to linear combinations of the lowest frequency modes of the full set of solutions of the Dirac equation (\ref{eq:7}).

\paragraph{}Indeed, we calculate the center of mass $\langle\theta\rangle_{\psi_0}$ (\ref{eq:3}) of the Dirac fermion associated to the spinor $\psi_0$ and find
\ba
\langle\th\rangle_{\psi_0}&=&A(\frac{Mcl}{\hbar})\,b\, \left[d \underbrace{\cos(\frac{ct}{l})}_{\mathrm{geodesic\, motion}} + 
			b\,\,\,\, \underbrace{\cos((\frac{c}{l} + 2 \frac{c^2M}{\hbar}) t)}_{\mathrm{Zitterbewegung}}\right]\label{psi}\ea
It is indeed a superposition of two oscillations that respectively describe its geodesic motion and the so-called Zitterbewegung. The respective frequencies are given by
		
        \begin{itemize}
            \item Geodesic motion:\be\omega_{\mathrm{Geo}}=\frac{c}{l},\ee

            \item Zitterbewegung: \be\o_{\mathrm{ZB}}=\frac{c}{l}+2\frac{c^2M}{\hbar}.\label{ZB}\ee
        \end{itemize}
    For these solutions $\pg{\psi_0}$, the Zitterbewegung frequency is inversely proportional to the curvature radius of $\mathrm{AdS}_2$ and proportional to the mass. In particular, we recover the flat space result when taking the flat space limit, $l\rightarrow\infty$ (in natural units)
    \be \o_{\rm{ZB}}=\frac{c}{l}+2\frac{c^2M}{\hbar}\rightarrow 2\frac{c^2M}{\hbar}.\label{corr}\ee

Finally, we compute the center of mass of the lightbeam $\langle n\rangle_{\psi_0}$ using the direct relation derived above (\ref{threln}). We obtain that
\ba
            \langle n\rangle_{\psi_0}&=&A_{\mathrm{Geo}}\cos(\frac{ct}{l}))+A_{\mathrm{ZB}}\cos((\frac{c}{l} + 2 \frac{c^2M}{\hbar}) t))+A_{\mathrm{corr}}\cos((\frac{2c}{l}+2\frac{Mc^2}{\hbar})t).\hspace{1cm}\label{cont}
\ea

\paragraph{} We find that it differs from the center of mass of the Dirac fermion $\langle\theta\rangle_{\psi_0}$ (\ref{psi}) by a third oscillation with frequency (in the natural units)
\be \o_{\mathrm{corr}}=\frac{2c}{l}+2\frac{c^2M}{\hbar}\label{corr2}.\ee
In this work, we consider the regime $\frac{Mcl}{\hbar}\gg1$ so that the
frequency of the correction, $\o_{\mathrm{corr}}$ (\ref{corr2}) is very close to the Zitterbewegung frequency $\o_{\mathrm{ZB}}$ (\ref{ZB}). The importance of this correction depends on the ratio of amplitudes$\frac{A_{\mathrm{Geo}}}{A_{\mathrm{corr}}}$ and $\frac{A_{\mathrm{corr}}}{A_{\mathrm{ZB}}}$ and is determined by
the initial conditions of the continuous spinor $\psi_0$ (\ref{sol}). To be consistent with the experiments, we would like the amplitude of the initial spinor $|\psi_0(t=0)|$ to be a Gaussian. However, it is given by a linear combination of 
$(\cos \th)^{2lM} \sin \th$, $(\cos \th)^{2lM}$ and $\cos \th^{2lM} \cos 2\th$ and such linear combinations do not result in an exact Gaussian. We can only choose the initial spinor $\psi_0(t=0)$ such that its amplitude is well approximated by a Gaussian. We call such initial state “Gaussian-like”.
By tuning $\psi_0(t=0)$ to be “Gaussian-like”, we find that $A_\mathrm{Geo} > A_\mathrm{ZB}, A_\mathrm{corr}$. Hence, the center of mass of the lightbeam $\langle n\rangle_s$ predicted by the numerical simulations (\ref{sim}) matches both with the center of mass of the lightbeam $\langle n\rangle_{\psi_0}$  (\ref{cont}) calculated from the continuous spinor $\psi_0$, and with the center of mass of the Dirac fermion $\langle\theta\rangle_{\psi_0}$ (\ref{psi}).

\printbibliography
%    \subsection{Conclusion}
%\paragraph{}We have shown how the propagation of Dirac fermions in a curved spacetime and in particular in a two dimensional Anti-de-Sitter spacetime $\mathrm{AdS}_2$ can be emulated using 1-dimensional waveguide lattices. This relies on the discretization of the Dirac equation \cite{Koke_2016} which allows to map the time evolution of Dirac fermions in $\mathrm{AdS}_2$ to the spatial propagation of light in a one dimensional waveguide array with a constant coupling parameter and a $\frac{1}{\cos\th}$ detuning parameter. In particular, the center of mass motion of the Dirac fermion is mapped to the center of mass motion of the light beam: the light undergoes a superposition of two periodic motions corresponding to the geodesic (or main) motion of the particle and the Zitterbewegung, a small oscillation around its main trajectory.

\end{document}